\documentclass[apj]{emulateapj}
\newcommand\submitms{n}		
\if\submitms y
\else
\usepackage{apjfonts}
\fi

\usepackage{graphicx}
\usepackage{ifthen}
\usepackage{natbib}

\shorttitle{SL9 Plume Development}
\shortauthors{Palotai {\em et al.}}

\newcommand\degree{\degr}
\newcommand\degrees\degree

\newcounter{fignum}

\DeclareSymbolFont{UPM}{U}{eur}{m}{n}
\DeclareMathSymbol{\umu}{0}{UPM}{"16}
\let\oldumu=\umu
\renewcommand\umu{\ifmmode\oldumu\else\math{\oldumu}\fi}
\newcommand\micro{\umu}
\renewcommand\micron{\micro m}
\newcommand\microns \micron

\let\oldsim=\sim
\renewcommand\sim{\ifmmode\oldsim\else\math{\oldsim}\fi}
\let\oldpm=\pm
\renewcommand\pm{\ifmmode\oldpm\else\math{\oldpm}\fi}
\newcommand\by{\ifmmode\times\else\math{\times}\fi}

\newcommand\tablebox[1]{\begin{tabular}[t]{@{}l@{}}#1\end{tabular}}
\newbox{\wdbox}
\renewcommand\c{\setbox\wdbox=\hbox{,}\hspace{\wd\wdbox}}
\renewcommand\i{\setbox\wdbox=\hbox{i}\hspace{\wd\wdbox}}

\newcommand\herenote[1]{{\bfseries #1}\typeout{======================> note on page \arabic{page} <====================}}

\newcount\timect
\newcount\hourct
\newcount\minct
\newcommand\now{\timect=\time \divide\timect by 60
         \hourct=\timect \multiply\hourct by 60
         \minct=\time \advance\minct by -\hourct
         \number\timect:\ifnum \minct < 10 0\fi\number\minct}

\catcode`@=11

\if\submitms y
\renewcommand\comment[1]{}
\else
\newcommand\comment[1]{}
\fi
\newcommand\commenton{\catcode`\%=14}
\newcommand\commentoff{\catcode`\%=12}

\renewcommand\math[1]{$#1$}
\newcommand\mathshifton{\catcode`\$=3}
\newcommand\mathshiftoff{\catcode`\$=12}

\comment{the backslash is necessary}

\comment{alignment tab}

\let\atab=&
\newcommand\atabon{\catcode`\&=4}
\newcommand\ataboff{\catcode`\&=12}

\let\oldmsp=\sp
\let\oldmsb=\sb
\renewcommand\sp[1]{\ifmmode
	   \oldmsp{#1}%
	 \else\strut\raise.85ex\hbox{\scriptsize #1}\fi}
\renewcommand\sb[1]{\ifmmode
	   \oldmsb{#1}%
	 \else\strut\raise-.54ex\hbox{\scriptsize #1}\fi}
\newcommand\msp[1]{\ifmmode
	   \oldmsp{#1}
	 \else \math{\oldmsp{#1}}\fi}
\newcommand\msb[1]{\ifmmode
	   \oldmsb{#1}
	 \else \math{\oldmsb{#1}}\fi}
\newcommand\supon{\catcode`\^=7}
\newcommand\supoff{\catcode`\^=12}
\newcommand\subon{\catcode`\_=8}
\newcommand\suboff{\catcode`\_=12}
\newcommand\supsubon{\supon \subon}
\newcommand\supsuboff{\supoff \suboff}

\newcommand\actcharon{\catcode`\~=13}
\newcommand\actcharoff{\catcode`\~=12}

\newcommand\paramon{\catcode`\#=6}
\newcommand\paramoff{\catcode`\#=12}

\comment{And now to turn us totally on and off...}

\newcommand\reservedcharson{\commenton \mathshifton \atabon \supsubon \actcharon
	\paramon}

\newcommand\reservedcharsoff{\commentoff \mathshiftoff \ataboff
	\supsuboff \actcharoff \paramoff}

\newcommand\nojoe[1]{\reservedcharson#1\reservedcharsoff}

\catcode`@=12
\reservedcharsoff

\reservedcharson

\comment{ Must have ONLY ONE of these... trust these macros, they work

}
\reservedcharsoff

\actcharon
\if\submitms y

\else

\received{2010 November 22.}
\comment{\revised{2011 January 16.}}
\accepted{2011 January 18.}
\fi

\if\submitms y
\slugcomment{\tablebox{
Submitted to {\em ApJ} 2010 November 18. Accepted 2011 January 18.}}
\else
\slugcomment{Version of 2011 Feb 15 for arXiv.org.}
\journalinfo{{\em To appear in} The Astrophysical Journal.}
\fi

\begin{document}

\title{PLUME DEVELOPMENT OF THE SHOEMAKER-LEVY 9 COMET IMPACT}
\author{Csaba Palotai\altaffilmark{1}}
\author{Donald G.\ Korycansky\altaffilmark{2}}
\author{Joseph Harrington\altaffilmark{1}}
\author{No\'emi Rebeli\altaffilmark{1}}
\author{Travis Gabriel\altaffilmark{1}}
\email{csaba@physics.ucf.edu}

\affil{\sp{1}Planetary Sciences Group, Department of Physics,
 University of Central Florida, Orlando, FL 32816-2385, USA}
\affil{\sp{2}Department of Earth and Planetary Sciences, University of California, Santa Cruz,
CA 95064, USA}

\begin{abstract}
\if\submitms y
\else
\comment{\hfill\herenote{DRAFT of {\today} \now}.}
\fi
We have studied plume formation after a Jovian comet impact using the
ZEUS-MP 2 hydrodynamics code.  The three-dimensional models followed
objects with 500, 750, and 1000 meter diameters.  
Our simulations show the development of a fast,
upward-moving component of the plume in the wake of the impacting
comet that ``pinches off'' from the bulk of the cometary material \sim50 km below the 1 bar pressure level, \sim100 km above the depth of greatest
mass and energy deposition.  The fast-moving component
contains about twice the mass of the initial comet, but consists
almost entirely (\math{>}99.9\%) of Jovian atmosphere rather than cometary
material.  The ejecta rise mainly along the impact trajectory, but
an additional vertical velocity component due to buoyancy 
establishes itself within seconds of impact, leading to an asymmetry
in the ejecta with respect 
to the entry trajectory.  The mass of the upward-moving
component follows a velocity distribution \math{M(>v)} approximately
proportional to \math{v\sp{-1.4}} (\math{v\sp{-1.6}} for the 750 m and 500
m cases) in the velocity range \math{0.1 < v
  < 10} km s\math{\sp{-1}}.

\end{abstract}

\section{INTRODUCTION}
\label{intro}
In July 1994, several fragments of comet Shoemaker-Levy 9 (SL9)
impacted Jupiter, giving a unique opportunity for direct observation
of a hypervelocity impact and its aftermath.  The scientific community labeled it as ``once in a lifetime event'' because at that point the estimates for the probability of a similar event taking place suggested that decades or centuries might pass before it happens again.  The 2009 July 19 impact discovered by A.\ Wesley (Hammel et al.\ 2010, S\'anchez-Lavega et al.\ 2010) and the 2010 June 3 impact discovered by C.\ Go and A.\ Wesley \citep{hueso:2010} indicate that Jupiter impacts happen much more frequently.

\citet{harr:2004} give an overview of the phenomenology common to all
the larger impacts.  In the short but energetic entry phase the
fragments entered Jupiter's atmosphere at \sim45\math{\sp{\circ}}S
latitude with an impact velocity of over 60 km s\math{\sp{-1}} and
an impact angle of about 45\math{\sp \circ} \citep{chodas:1996}.
During entry, the impactors broke up and evaporated in 
\sim10 seconds \citep{kory:2006},
depositing most of their kinetic energy close to the terminal depth.
Each impact created 
a low-density entry channel consisting of high-temperature Jovian air
and impactor material
\citep{maclow:1994, zahnle:1994, zahnle:1995, crawford:1995,
  kory:2006}.  This
column of debris rose back up in the entry channel and expanded
radially, generating shock waves.  At higher altitudes, the plume rose
ballistically, with the visible top of the debris reaching about 3000
km above the cloud tops \citep{hammel:1995, jessup:2000}.  The plume
then collapsed, compressing and heating itself
and the upper atmosphere it encountered as it continued to expanded
radially \citep{deming:2001}.

Modeling the observed phenomena is difficult because the
various stages of the process have time and length scales that differ
by several orders of magnitude and the relevant physics and chemistry
cannot be incorporated into a single model.  Instead, modelers have
simulated individual phases of the 
impact.  Initial conditions for models of later phases (entry response, plume flight, plume
splash) come from remapping the final state of the previous
phase onto a larger grid.  \citet{zahnle:1994} and
\citet{crawford:1995} chained the entry and the blowout phases in this
manner, while \citet{harr:2001} and \citet{deming:2001} connected the
plume flight and landing response phases using the same technique in a
two-dimensional model.  Exceptions to this practice are
\citet{takata:1994} and \citet{takata:1997}, who used a smoothed
particle hydrodynamics (SPH) model to include both the entry and the
entry respond phases in their three-dimensional (3D)
simulations.  \citet{harr:2004} give a more complete review of
previous modeling efforts.

Since all subsequent phases depend on the entry,
\citep{kory:2006} performed high-resolution, 3D simulations, computing
energy deposition profiles and terminal depths
of various impactor types.  In this paper, our investigation focuses on
plume development in the immediate aftermath of an SL9 comet fragment's
entry.

\if\submitms y
\clearpage
\fi
\begin{figure*}[tb]
\if\submitms y 
  \setcounter{fignum}{\value{figure}}
  \addtocounter{fignum}{1} 
  \newcommand\fignam{f\arabic{fignum}.eps}
\else 
  \newcommand\fignam{f1.png} 
\fi
\strut\hfill
\includegraphics[width=\textwidth, clip]{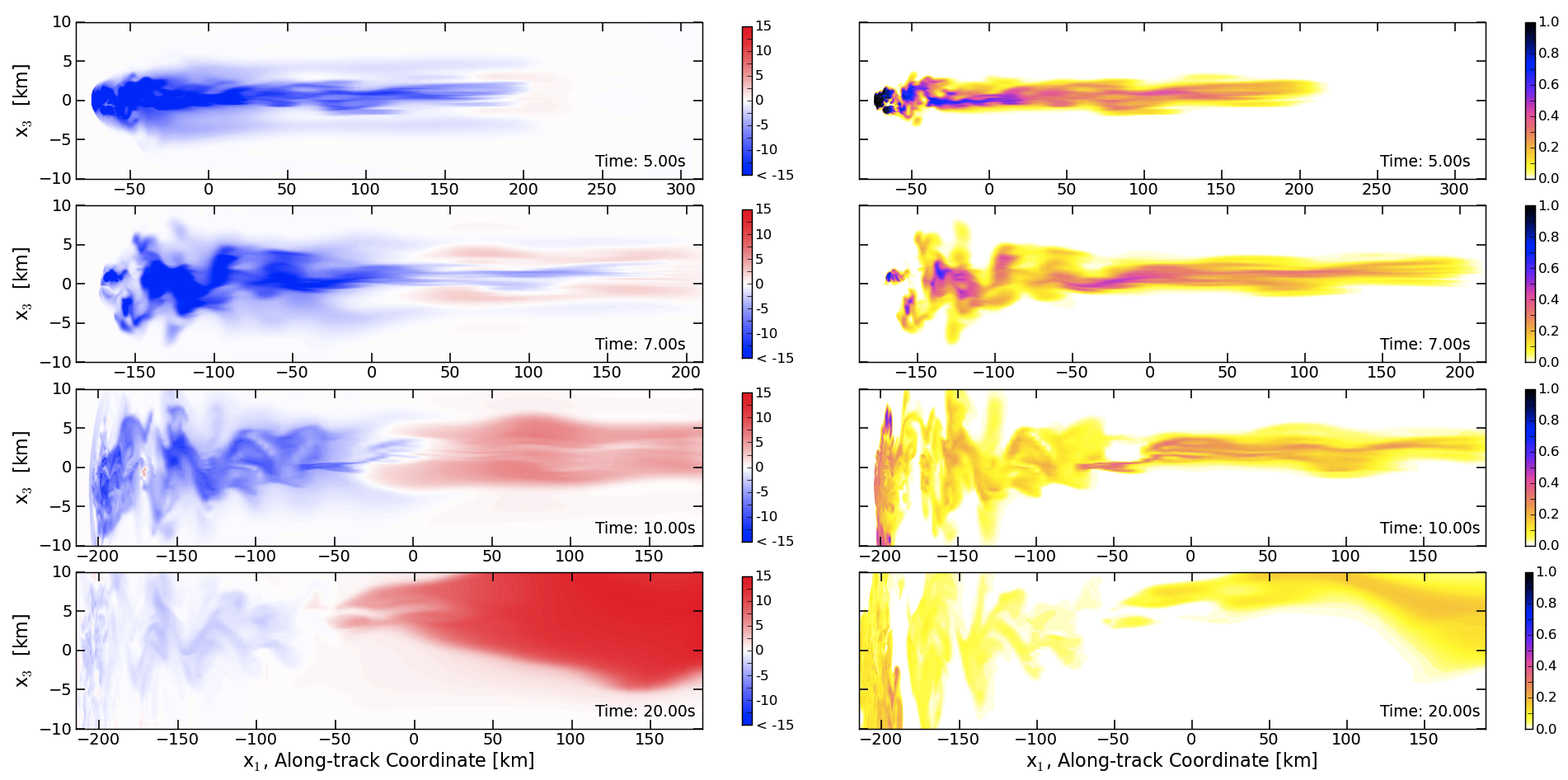}
\hfill\strut
\figcaption{\label{fig:plume} Plume genesis. {\bf Left:}
  Along-track velocity values from a 1 km diameter SL9 impact
  simulation.  Negative values represent material moving down.  The
  maximum velocity in the mature plume agrees well with the estimated
  ejection velocity of \citet{hammel:1995}.  {\bf Right:} Mass
  fraction of impactor material {\em vs.} Jovian air in the same
  simulation.  Note the position of the origin.  
  }
\end{figure*}
\if\submitms y
\clearpage
\fi

\section{MODEL DESCRIPTION}
\label{sec:model}
We modified the ZEUS-MP 2
hydrodynamics code \citep{hayes:2006} for Jovian impacts, validating
the modifications modifications with a series of tests.  These included
ZEUS-MP's own hydrodynamic test suite (e.g.\ Sedov-Taylor blast wave, radiating shock waves) and our simulations matched their published data \citep{hayes:2006}.  We also set up simulations to verify hydrostatic equilibrium and the long-term stability of the atmosphere
over many sound-crossing times.  Past work on atmospheric impacts 
\citep{kory:2002} has found significant differences in results between 2D and 3D simulations.  In particular geometrical constraints that operate in two dimensions (but not in three) tend to enforce global enstrophy conservation, which in turn forces a portion of the flow kinetic  energy into large scales.  Two-dimensional calculations thus show unrealistic amounts of large-scale structure \citep{khokhlov:1994}.  Because of the qualitative differences between 2D and 3D fluid dynamics, 3D models provide more accurate results than those of the SL9 era. 

In order to delay the remapping of variables to a
second model as long as possible, we studied the plume's genesis and
its effect on the Jovian atmosphere by extending the grid of
\citet{kory:2006} and continuing calculations beyond the \sim10
seconds of impactor destruction.  The sensitivity tests of
\citet{kory:2006} determined that having a higher resolution than 16
grid elements within the impactor's radius (R16) did not result in
significant changes in the impactor's disruption mode or penetration
depth, so we used R16 for the present models.
The Cartesian coordinate system aligns
with the initial trajectory of the impactor; \math{x\sb1} is the
along-track coordinate, \math{x\sb2} is horizontal, and \math{x\sb3} is
perpendicular to the others.  The 1 bar pressure level coincides with
the origin of the {\emph x\sb1} coordinate.  Away from the high-resolution
inner region that contains the impactor, the spacing increases by
\sim4\% per grid cell.  A typical grid has a size of \sim
400\math{\times}90\math{\times}90 km and has \sim
430\math{\times}280\math{\times}280 grid elements.  These numbers vary
somewhat depending on the diameter of the impactor.  The
grid moves with the impactor to keep it in the high-resolution region,
and stops when the comet is vaporized.  
When the grid stops, there is no undisturbed Jovian air along the track of
the impactor and behind it.  This allows us to model the acceleration and expansion
of the plume as accurately as possible.  \citet{kory:2006} provide a
more-detailed description of the model.

\section{RESULTS}
\label{sec:results}
Our nominal case is that of \citet{kory:2006}, a 1 km diameter, 0.6 g
cm\math{\sp{-3}}, porous, spherical, ice impactor arriving at
44\math{\sp{\circ}}.02 S, at 61.46 km s\math{\sp{-1}}, at a
42\math{\sp{\circ}}.09 impact angle.  This case is
likely larger than a typical SL9 fragment \citep{carlson:1997} but it
lets us compare our results to prior work.

\begin{figure}[tb]
\if\submitms y 
  \setcounter{fignum}{\value{figure}}
  \addtocounter{fignum}{1} 
  \newcommand\fignam{f\arabic{fignum}.eps}
\else 
  \newcommand\fignam{f2.png} 
\fi
\strut\hfill
\includegraphics[width=\columnwidth, clip]{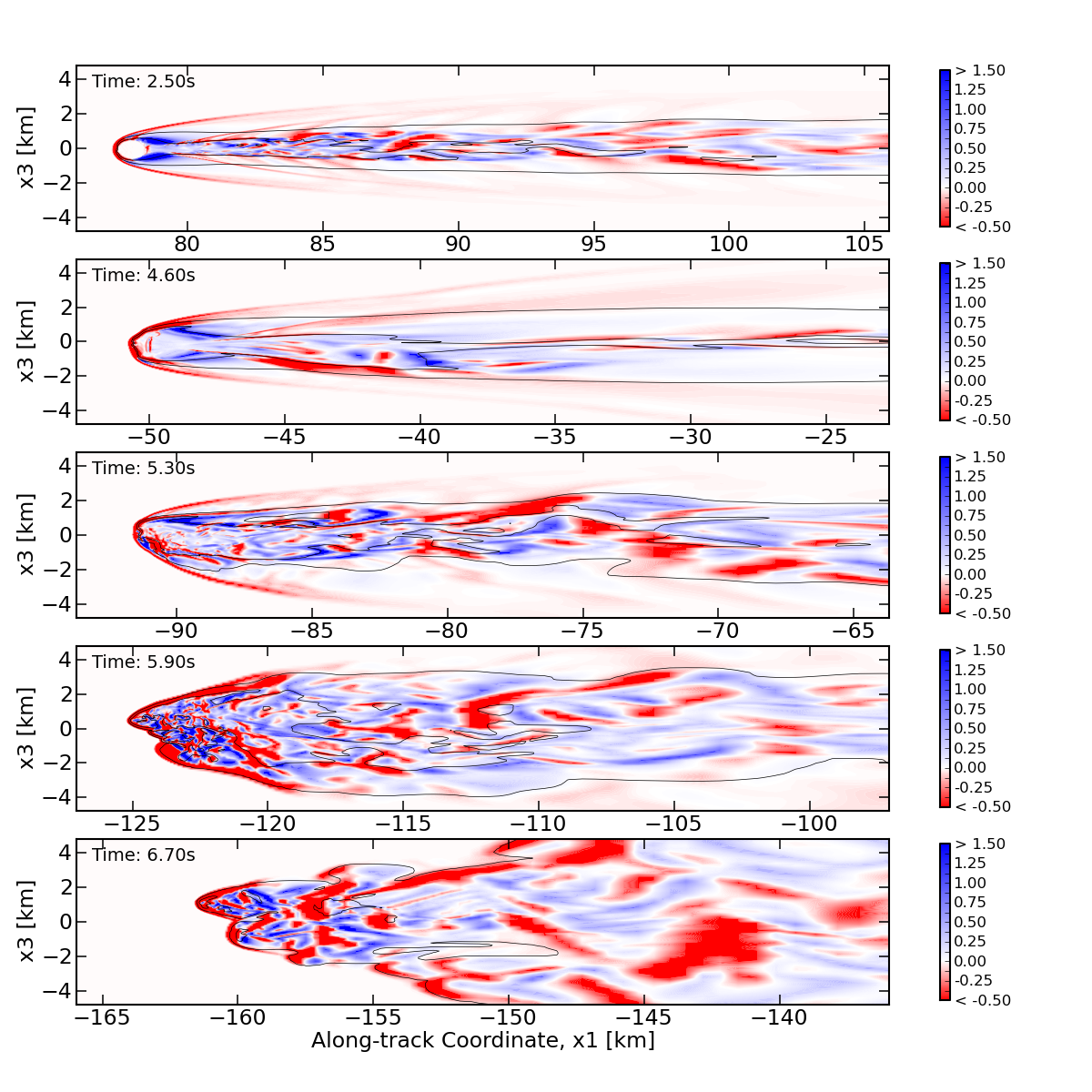}
\hfill\strut
\figcaption{\label{fig:shock} The shock structure and the distribution of impactor debris during the breakup an SL9-type impactor.  The colot
  variable is \math{dv/dx~ [10\sp{9}} s\math{\sp{-1}}]. The purpose of the color is not to emphasize the variable itself but to indicate the location and the structure of the shocks.  Black contour lines represent the mass fraction of comet material, the outermost line represents 1\% of impactor material and the inner lines represents 80\%. Prior to the breakup, the shock system on the trailing edge of the impactor confines the debris that falls off of the impactor resulting in a narrow trail of cometary material within the plume.  After the breakup, which occurs at about 4.2 s, the shock system falls apart and allows for the spreading of the debris perpendicular to the entry path at greater depths.}
\end{figure}

Figure~\ref{fig:plume}. shows the impact and genesis of the plume.
Initially, the shock structure behind the impactor (Figure~\ref{fig:shock}.)
limits the distribution of cometary material to the turbulent
wake, so only a narrow trail of impactor debris remains in the entry
channel.  The breakup of the impactor begins at about 4.2 s into the
simulation and at about 40--50 km below the 1 bar pressure level
(distances indicate along-track values).  This event disrupts the
organized flow behind the main fragment, allowing for the spreading of impactor material perpendicular to the entry path and more vigorous mixing of debris and Jovian air as the impactor descends to greater depths.

The plume forms as upward velocities appear within the
entry channel.  At this point we find an interesting and important
property of the plume: it ``pinches off'' at a certain depth.
Below this level, the material rises independently and much slower.  Almost all of the
cometary material is below this level, and although most of the energy
gets deposited near the terminal depth, this material only rises
buoyantly.  Most likely it cannot achieve ballistic ejection (see
below).  We associate the pinch-off level with the breakup level of the
impactor and the disruption of the simple entry shock structure.  The
500 m and 750 m 
impactors pinch off within about a scale height of the nominal
case.

Based on their 2D experiments, \citet{boslough:1995} reported that
using only the energy deposited by a 3 km diameter impactor above the
-50 km level results in an almost identical fireball to that generated
by using the full energy deposition curve.  They concluded that the
fireball growth depends mostly on the diameter of the impactor and the
fact that several plumes reached the same height \citep{hammel:1995}
implies that the fragments that generated those plumes were the same
size.  In our simulations the plume detachment occurs at about the
same depth that \citet{boslough:1995} used for the energy deposition
cutoff, which strengthens the hypothesis that the energy deposition
below this level (more than 99\% of the total energy deposition) would
not affect the buildup of the plume.  However, in our models the 1 km,
750 m, and 500 m diameter impactors all broke up at similar altitudes
and all of them generated ejection velocities of \math{>}12 km
s\math{\sp{-1}} despite the factor of eight range of mass/kinetic
energy and factor of four range in initial cross-section.  Thus, the
observed plume heights likely could have been achieved by
different-sized impactors.  We plan future studies to verify this.

\begin{figure}[tb]
\if\submitms y 
  \setcounter{fignum}{\value{figure}}
  \addtocounter{fignum}{1} 
  \newcommand\fignam{f\arabic{fignum}.eps}
\else 
  \newcommand\fignam{f3.png} 
\fi
\strut\hfill
\includegraphics[width=\columnwidth, clip]{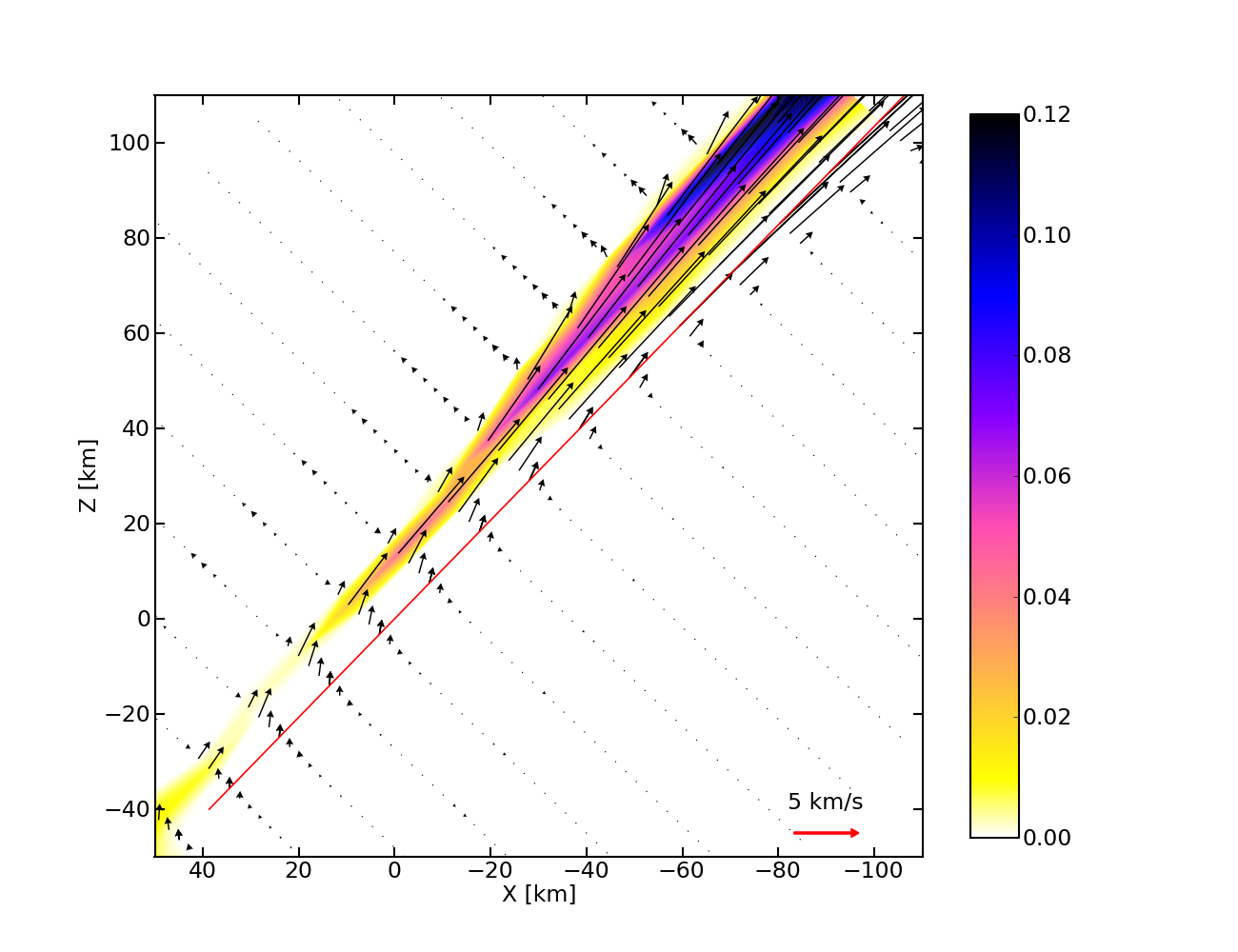}
\hfill\strut
\figcaption{\label{fig:vector} Velocity distribution in our 1 km
  diameter impactor model, shown in ``true'' vertical coordinates.
  The red line indicates the entry trajectory.  Color shows the
  impactor mass fraction in the ejecta.}
\end{figure}

\citet{crawford:1995} noted that during the early stages of their
model the evolving fireball maintained axisymmetry with respect to the
entry channel.  The oblique impact simulations of
\citet{boslough:1995} and \citet{takata:1994}
also predict the ejecta to rush back along the entry
trajectory.  In their models, \citet{pankine:1999}
aligned the axis of the ejecta cone with the
trajectory of the incoming comet.  \citet{hammel:1995} estimated that
in order for the plume material to reach the observed height of 3000
km at the 45\math{\sp{\circ}} ejection angle, an initial velocity of
\sim17 km s\math{\sp{-1}} is required for ballistic particles.
However, \citet{harr:2001} found that the best fit to the appearance
of the impact site came with an ejection cone tilted just 30{\degrees}
from the vertical.
Our models do not produce velocities that high even for the 1 km
diameter impactor, although it is conceivable that they may
  appear later as the plume blows out from the atmosphere.  However, even in the
first few seconds of plume formation, 
the entry channel expands mostly toward low pressure, creating an
asymmetry with respect to the entry channel.  Figure \ref{fig:vector}
shows the debris ejection directions in our
model.  Most of the ejecta are above the impactor's trajectory,
and the deviation from the entry path increases with time.
\citet{jessup:2000} determined that an initial tilt angle of
\sim6\math{\sp{\circ}}--15\math{\sp{\circ}} and initial velocity of
``only'' \sim 11--12 km s\math{\sp{-1}} are required to reproduce the
time-dependent behavior of the observed plume apexes from the A, E, G,
and W impacts.  We do see velocities that high in our model, even in
the cases of the 750 m and 500 m diameter impactors.

\if\submitms y
\clearpage
\fi
\if\submitms y
\clearpage
\fi

\citet{jessup:2000} deduced from the Galileo Near-Infrared Mapping Spectrometer
data that the G fireball was most likely
initiated between the 100 and 200 mbar pressure levels (35--45 km above
the 1 bar level).  Figure~\ref{fig:plume} shows that our model agrees
with this deduction: at this level (\sim50--70 km in along-track
coordinates) the ejection velocities are near their maxima.

\if\submitms y
\clearpage
\fi
\if\submitms y
\clearpage
\fi

\citet{takata:1994} modeled the evolution of the plume expansion.
They observed a gradual upward ejection of the debris and estimated
that more than 40\% of cometary material would rise above the 1 bar
level at 100 s into their simulation, and more than 80\% would
eventually ascend above the visible cloud decks.
\citet{crawford:1995} calculated that the energy and mass deposition
of large fragments (\math{>}1 km diameter) occurs below Jupiter's ammonia
cloud layer and predicted that less than 1\% of the impactor material
would be entrained into the rising plume.  Lagrangian tracer particles
in their model indicated that for a 3 km diameter impactor the total
material in the plume above the 1 bar level is about 6.8 fragment
masses, with only 0.2\% being cometary material.  \citet{carlson:1997}
modeled the mass outflow of the G-impact.  They assumed equal Jovian
and cometary contributions in the plume, with a total mass of \sim
\math{2.2 \times 10\sp{13}} g.  Based on the chemistry model of
\citet{zahnle:1995b}, they assumed that 40\% of the comet material
would produce water, which agrees with the observations by
\citet{bjoraker:1996} and \citet{encrenaz:1997} for the G plume
splash.  At 20 s into our simulation, the plume is completely
detached from the rest of the comet-disturbed material and the mass of
the debris that is moving with an upward velocity component of more
than 100 m s\math{\sp{-1}} is about \math{1.6 \times 10\sp{15}} g,
more than three times the mass of the original 1 km diameter fragment.
The material with cometary origin in the plume at the same time is
about \math{4.0 \times 10\sp{11}} g, about 0.07\% of the original
fragment or about 0.02\% of the total plume material.  For the 750 m
and 500 m diameter impactors the plume masses are \math{{1.1} \times
  10\sp{15}} g and \math{{3.4} \times 10\sp{14}} g, and the comet
fractions are \math{{1.4} \times 10\sp{11}} g and \math{{9.5} \times
  10\sp{10}} g, respectively.

\citet{boslough:1995} introduce an upwelling phase.  At the depth of
maximum energy deposition, a bubble of material forms from cometary
material and Jovian air and rises buoyantly.  Based on simulations of
\citet{crawford:1995}, they suggested that the adiabatic expansion of
this bubble as it approaches the visible cloud deck is the source for
the expanding ring of \citet{hammel:1995}.  In our 80 s simulations
(not shown), we observe a similar feature forming out of the material
below the pinch-off level.  We plan experiments to
study whether this feature, plume-related shocks, or something
else forms the expanding ring.  Our simulations indicate that the blowout plume should be relatively water free and most likely it cannot serve as the source of the observed water abundance values \citep{bjoraker:1996,encrenaz:1997}. Alternatively, this potentially water-rich bubble that forms well below the water cloud level and contains most of the comet material could provide an explanation for those observations.  There is also the possibility of water forming from other species in the stratosphere and detailed investigation of these hypotheses is needed.

\if\submitms y
\clearpage
\fi
\begin{figure}[tb]
\if\submitms y 
  \setcounter{fignum}{\value{figure}}
  \addtocounter{fignum}{1} 
  \newcommand\fignam{f\arabic{fignum}.eps}
\else 
  \newcommand\fignam{f4.png} 
\fi
\strut\hfill
\includegraphics[width=\columnwidth, clip]{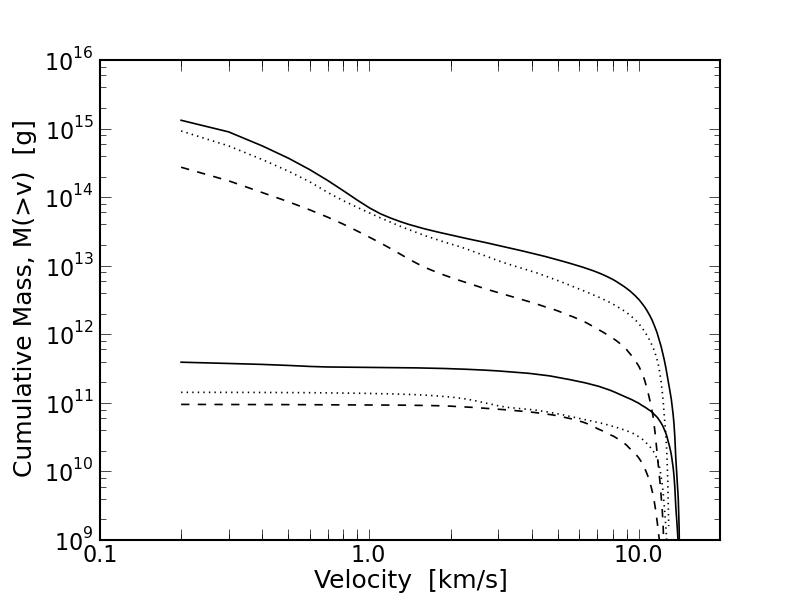}
\hfill\strut
\figcaption{\label{fig:cmass} The mass-velocity distribution for 1000
  m (solid), 750 m (dotted), and 500 m (dashed) diameter impactors at
  20 s.  The top curves are for total upward-moving mass; the
  bottom curves are for upward-moving comet material only.}
\end{figure}
\if\submitms y
\clearpage
\fi

The horizontal (\math{x\sb1-x\sb2}-plane) opening angle of the plume cone
is \sim25\math{\sp{\circ}} at the end of our 25--30 s
simulations.  This value is lower than expected based on the size of the
crescent-shaped debris field.  \citet{harr:2001} used a
70\math{\sp{\circ}} opening angle to model the crescent and
\citet{pankine:1999} also stated that the opening angle should be
about 70\math{\sp{\circ}} to produce a crescent that spans
180\math{\sp{\circ}} around the impact site.  The smaller opening angle in our model is the result of the plume leaving the computational grid, but we anticipate that the further expansion of the plume at higher altitudes will allow it to open more and we will model this by re-mapping the variables onto a larger grid.  At 30 s in our simulation, the small amount of cometary material is limited horizontally to a very narrow trail within the plume.  This could lead to a higher abundance of cometary material in the middle of the crescent than towards the sides.  We plan to study this in the future.

\citet{zahnle:1995} calculated mass-velocity distributions (MVD) from
simulations similar to those of \citet{zahnle:1994}.  The MVD of
ejecta from hypervelocity impacts usually follows a power law that can
be written in a cumulative form
\begin{eqnarray*}
\label{eq:cmvd}
M(>v) \propto v\sp{-\alpha},
\end{eqnarray*}
where \math{M(>v)} is the mass at velocities
greater than \math{v}.  Their power-law exponent was \math{\alpha =
  1.55}.  The conservation laws of energy and momentum constrain
\math{\alpha} to be \math{1.0 < \alpha < 2.0}.  \citet{harr:2001} used
this distribution in their ballistic Monte Carlo plume model.
\citet{pankine:1999} used two different MVDs in order to simulate the
plume and the resulting debris field.  The first distribution was
based on isentropic expansion of gas into vacuum following an
explosion.  The second MVD assumed that all the mass was ejected at
the same velocity.

Figure \ref{fig:cmass} shows the MVDs from our simulations.  The
results follow power law curves below and above the cutoff velocity of
\sim10 km s\math{\sp{-1}}.  The plume
generated by the 1 km diameter impactor has \math{\alpha} \sim 1.4; for
the 750 m and 500 m diameter cases \math{\alpha} \sim 1.6.  For the comet material only (bottom
curves), these curves are much shallower, indicating that proportionally more
impactor material ejects at higher speeds.

\section{CONCLUSIONS}
\label{sec:disc}
In this study we focused on the generation and evolution of the
initial plume after a Jovian cometary impact.
We found that there is a pinch-off point approximately 50 km below the
1 bar level in the atmosphere, above which the initial plume begins to
form and ascend, with velocities up to \sim15 km s\math{\sp{-1}}. The
initial opening angle of the plume is quite narrow, much smaller than
the 70\math{\sp{\circ}} found in other models.  However, further
evolution of the plume as it blows out may change this value.  The
bulk of the cometary material and kinetic energy penetrates to much
lower depths; it rises buoyantly and
much slower, and merits our further study.  The fast plume contains
more than three times the mass of the original
impactor but is almost entirely (\math{>}99.9\%) composed of material from
the Jovian atmosphere, so the observed dark features away from the
impact site (e.g., main
ring, crescent) should contain almost exclusively processed Jovian air.
We will continue to investigate Jovian impacts with the
goal of relating their physical properties to the observations.

\acknowledgments This research was supported by National Science
Foundation Grants AST-0813194 and AST-0964078.

\nojoe{
\if\submitms y
  \newcommand\bblnam{ms}
\else
  \newcommand\bblnam{apj2010plume}
\fi
\bibliography{\bblnam}}

\end{document}